\documentstyle[prl,aps,tighten,epsf,epsfig]{revtex}
\newcommand{\drv}[2]{\frac{d#1}{d#2}}
\def\stacking#1#2#3{\mathrel{\mathop{#3}\limits^{#1}\limits_{#2}}}

\begin{document}


\newpage
\title{A theory for the membrane potential of cells.}
\author{Lars Petter Endresen$\, \dagger$ and Kevin Hall$\, \ddagger$}
\address{$\dagger \,$Institutt for fysikk, NTNU, N-7034 Trondheim, Norway and
         $\ddagger \,$Centre for Nonlinear Dynamics, McGill University, Montreal, Canada}
\date{\today}
\maketitle

\section*{Abstract}
\begin{abstract}
We give an explicit formula for the membrane potential of cells in
terms of the intracellular and extracellular ionic concentrations, and
derive equations for the ionic currents that flow through channels,
exchangers and electrogenic pumps based on simple energy
considerations and conservation laws. We demonstrate that the work
done by the pump is equal to the potential energy of the cell plus the
energy loss due to the downhill ionic fluxes through the channels and
the exchanger. Our equations predict osmotic pressure variations. The
theory is illustrated in a simple model of spontaneously active cells
in the cardiac pacemaker. The simulated action potential and the five
currents in the model are in excellent agreement with experiments. The
model predicts the experimental observed intracellular ionic
concentration of potassium, calcium and sodium. We do not see any
drift of the values for the concentrations in a long time simulation,
instead we can obtain the same asymptotic values starting with equal
intracellular and extracellular ionic concentrations.
\end{abstract}

\section{Introduction}
Ionic models of the cellular membrane potential are typically guided
by extensive data sets from voltage clamp experiments. The purpose of
this paper is not to introduce a new ionic model incorporating the
many experimentally observed currents (Boyett, 1996), or to critically
review other models, but merely to give an alternative formulation
based on simple principles from classical physics. Instead of trying
to reproduce observed voltage clamp data and tail currents of
activation, deactivation and inactivation, we demonstrate that simple
energy considerations and conservation laws can be useful when
formulating a model.

We derive equations for ionic currents flowing through channels,
exchangers and electrogenic pumps. These are based on the Boltzmann
distribution law (Boltzmann, 1868), which states that a particle in
thermal equilibrium spends less time in states of higher energy than
in states of lower energy, the Markov assumption (Markov, 1906) which
says that the transition probabilities to the next state of a
stochastic system is only dependent on its present state, and the
principle of detailed balance (Onsager, 1931) which says that the
microscopic laws of physics are invariant with respect to the reversal
of time. Our equations were inspired by Ehrenstein and Lecar's model
of channel gating (1977), Mullins' model of the ${\rm Na}^{+},{\rm
Ca}^{2+}$ exchanger (1977), and Chapman's model of the ${\rm
Na}^{+},{\rm K}^{+}$ pump (1978).

Since the intracellular ionic concentrations are treated as dynamical
variables, we show that the standard differential equation for the
membrane potential can be replaced with a new algebraic equation. We
demonstrate that the pump generates a transmembrane voltage gradient,
a transmembrane pressure gradient, transmembrane concentration
gradients, and downhill ionic fluxes through the exchanger and the
channels.

The theory is illustrated with a simple model of spontaneously active
cells in the rabbit sinoatrial node. The observable parameters in the
model are based on the experiments of Shibasaki (1987), Hagiwara {\em
et al.} (1988), Muramatsu {\em et al.} (1996) and Sakai {\em et al.}
(1996). The non--observable parameters in the model are determined
numerically, in the same way as in an earlier study (Endresen, 1997a),
by comparing the action potentials generated by the model with the
action potentials recorded by Baruscotti {\em et al.} (1996).

\section{Derivation of the Equations}
\subsection{Nernst Equation}
There are two basic principles behind the average motion of
particles. The first is diffusion which is general; the second applies
only to charged particles such as ions in solutions. Simple diffusion
is described by the empirical law of Fick:
\begin{equation}
\label{eq1}
\vec{\phi} = - ukT \nabla [{\rm S}] \;,
\end{equation}

\noindent
where ${\phi}$ is the ionic flux, $[{\rm S}]$ the concentration of ions
and $u$ the ratio of the velocity to the force acting on a particle,
known as the mobility. The empirical law of Ohm describes the net
motion of charged particles in an electric field:
\begin{equation}
\label{eq2}
\vec{\phi} = - zeu [{\rm S}] \nabla v \;,
\end{equation}

\noindent
where $z$ is the valence, $e$ the elementary charge and $v$ the
potential. Since we assume that the temperature is constant, we can
neglect the thermal flux given by Fourier's empirical law. The fact
that the mobility in Fick's law must be identical to the mobility in
Ohm's law was first noticed by Einstein (1905). If we combine
equations (\ref{eq1}) and (\ref{eq2}), the total flux of ions due to
diffusion and electric forces becomes:
\begin{equation}
\label{eq3}
\vec{\phi} = - ukT \exp\left(-\frac{zev}{kT}\right) \nabla \left[[{\rm S}] \exp\left(\frac{zev}{kT}\right)\right] \;.
\end{equation}

\noindent
The equilibrium potential where the flux is zero can be found by
integrating equation (\ref{eq3}) from the intracellular (i) to the
extracellular (e) side of the membrane:
\begin{equation}
\label{eq6}
v_{\rm S} = v_{\rm i}-v_{\rm e} =  \frac{kT}{ze} \ln \frac{[{\rm S}]_{\rm e}}{[{
\rm S}]_{\rm i}}\;,
\end{equation}

\noindent
where $[{\rm S}]_{\rm i}$ and $[{\rm S}]_{\rm e}$ refer to the
intracellular and extracellular concentrations. This equation, first
stated by Nernst (1888) is based only on the empirical laws of Ohm and
Fick and the relation of Einstein.

The same formula can be derived in a more general way using the
Boltzmann factor (Boltzmann, 1868). The relative probability at
equilibrium that an ion is at the intracellular or extracellular side
of a cell membrane is
\begin{equation}
\label{eq8}
\frac{p_{\rm e}}{p_{\rm i}} = \frac{[{\rm S}]_{\rm e}}{[{\rm S}]_{\rm i}} = \exp\left(-\frac{\Delta w}{kT}\right) \;,
\end{equation}

\noindent
where $\Delta w$ is the energy difference between these intra-- and
extracellular states , $T$ is the absolute temperature and $k$ is
Boltzmann's constant. If we consider ions of valence $z$ the energy
difference between the intracellular and extracellular side can be
written:
\begin{equation}
\label{eq9}
\Delta w = ze(v_{\rm e} - v_{\rm i}) \;.
\end{equation}

\noindent
This together with equation (\ref{eq8}) yields equation
(\ref{eq6}). The equilibrium potentials for the predominant cellular
cations are then:
\begin{eqnarray}
\label{eq11}
v_{\rm K}  &=& \frac{kT}{e} \ln \frac{[{\rm K}]_{\rm e}}{[{\rm K}]_{\rm i}}\;, \\
\label{eq12}
v_{\rm Ca} &=& \frac{kT}{2e} \ln \frac{[{\rm Ca}]_{\rm e}}{[{\rm Ca}]_{\rm i}}\;, \\
\label{eq13}
v_{\rm Na} &=& \frac{kT}{e} \ln \frac{[{\rm Na}]_{\rm e}}{[{\rm Na}]_{\rm i}}\;.
\end{eqnarray}

\subsection{Ionic Channels}
We assume that ionic channels are either completely open or completely closed and 
randomly fluctuate between these states. Each channel is here assumed to be a
simple Markov process (Markov, 1906), described by first order kinetics::
\begin{equation}
\label{eq25}
\begin{array}{ccc}
{\displaystyle C} & \stacking{\alpha}{\beta}{\displaystyle \rightleftharpoons} & {\displaystyle O}
\end{array} \;,
\end{equation}
where the rate constants ${\alpha}$ and ${\beta}$ are functions of
transmembrane voltage and control the transitions between the closed
($C$) and the open ($O$) states of the channel. The rate for a closed
channel to open is ${\alpha}$, and ${\beta}$ is the rate for an open
channel to close. Let $x$ denote the average fraction of channels that
are open, or, equivalently, the probability that a given channel will
be open. We may say that the ionic flux through an ensemble of
channels is regulated by a sliding door whose position is $x$. This
yields:
\begin{equation}
\label{eq26}
\drv{x}{t} = \alpha (1-x) - \beta x \equiv \frac{x_{\infty} -x}{{\tau}} \;,
\end{equation}
where
\begin{eqnarray}
\label{eq27}
x_{\infty} &=& \frac{{\alpha}}{{\alpha} + {\beta}} \\
\label{eq28}
\tau   &=& \frac{1}{{\alpha} + {\beta}}\;.
\end{eqnarray}

\noindent
Here $x_{\infty}$ denotes the steady state fraction of open channels
and ${\tau}$ the relaxation time. Let us assume that the energy
difference between the closed and open positions is given by

\begin{equation}
\label{eq29}
\Delta w =  w_{\rm open} - w_{\rm closed} \equiv q (v_{\rm x}-v)\;,
\end{equation}

\noindent
where $q$ is an gating charge, usually $q \approx 4e$, the term
$qv_{\rm x}$ is due to the difference in mechanical conformational
energy between the two states and $qv$ represents the change in
electrical potential energy due to the redistribution of charge during
the transition. At equilibrium, $dx/dt=0$, and the ratio of the
probabilities for a single channel to be in the open state or the
closed state is,
\begin{equation}
\label{eq30}
\frac{x_{\infty}}{1-x_{\infty}} = \frac{{\alpha}}{{\beta}}\;.
\end{equation}

\noindent
This relation is known as the principle of detailed balance (Onsager,
1931). The same ratio is given by the Boltzmann distribution
(Boltzmann, 1868),
\begin{equation}
\label{eq31}
\frac{x_{\infty}}{1-x_{\infty}} = \exp\left(-\frac{\Delta w}{kT}\right) \;.
\end{equation}

\noindent
Thus, from equations (\ref{eq29}), (\ref{eq30}) and (\ref{eq31}):

\begin{equation}
\label{eq31b}
{x}_{\infty} = \left( {1+\exp\left[\-\frac{(v_{\rm x}-v)}{\frac{kT}{4e}}\right]}\right)^{-1} \;,
\end{equation}

\noindent
and

\begin{equation}
\label{eq32}
\frac{\alpha}{\beta} = \frac{\exp\left[-\frac{(v_{\rm x}-v)}{\frac{kT}{2e}}\right]}
{\exp\left[+\frac{(v_{\rm x}-v)}{\frac{kT}{2e}}\right]} \;.
\end{equation}

\noindent
Therefore, making the {\em ad hoc} assumption that the product of
forward and backward reaction rates is constant, the symmetric choice
for $\alpha$ and $\beta$ is:

\begin{eqnarray}
\label{eq33}
{\alpha} &=& \lambda \exp\left[-\frac{(v_{\rm x}-v)}{\frac{kT}{2e}}\right] \\
\label{eq34}
{\beta} &=& \lambda \exp\left[+\frac{(v_{\rm x}-v)}{\frac{kT}{2e}}\right]\;,
\end{eqnarray}

\noindent
\noindent
where $\lambda$ is a constant. Taking $\lambda$ to be constant gives
the maximum relaxation time at the voltage where $x_{\infty}=1/2$. The
relaxation time is then:

\begin{equation}
\label{eq35}
{\tau} = \frac{1}{{\alpha}+{\beta}} = {\left(2\lambda \cosh\left[\frac{q(V_{\rm 1}-V)}{2kT}\right] \right)}^{-1}\;.
\end{equation}

\subsubsection{Potassium channels}

\noindent
If the flux of ions is driven by Ohm's law and regulated by the fraction of open channels 
$x$, the membrane current through potassium channels becomes:

\parbox{14cm}{
\begin{eqnarray*} 
i_{\rm K}    &=& g_{\rm K}\,x\left (v-v_{\rm K}\right ) \\
\frac{dx}{dt} &=&  {\tau}^{-1}_{\rm K} \cosh\left\{{\frac {v-v_{\rm x}}{\frac{kT}{2e}}}\right\}\left[ \frac{1}{2}\left (1+ \tanh\left\{{\frac { v-v_{\rm x}}{\frac{kT}{2e}}}\right\}\right )-x\right ] \;,
\end{eqnarray*}} \hfill
\parbox{1cm}{\begin{eqnarray}\label{eq36}\end{eqnarray}}

\noindent
where ${\tau}_{\rm K} = 1/{2\lambda}$ is the maximum value of the
relaxation time, $g_{\rm K}$ is the conductance, $v_{\rm K}$ is given
by equation (\ref{eq11}), where we have used the mathematical
identity:
\begin{equation}
\frac{1}{2} \left[1+ \tanh(\phi)\right] = \frac{1}{1+\exp(-2\phi)} \;.
\end{equation}

\subsubsection{Calcium and sodium channels}

\noindent
The calcium and sodium channels also have an inactivation mechanism 
in addition to the above activation mechanism. We can view these
mechanisms as two independent Markov processes, or equivalently two
independent sliding doors, which are both affected by voltage. An ion
can only go through if both sliding doors are at least slightly
open. The activation mechanism is very fast, so we use the steady
state fraction of open channels (equation (\ref{eq32})) for this. The
maximum time constant of inactivation for calcium and sodium 
channels are of the same order of magnitude as the maximum time
constant of the activation of the potassium channel, thus:

\parbox{14cm}{
\begin{eqnarray*}
i_{\rm Ca} &=& g_{\rm Ca}\,f\left (v-v_{\rm Ca}\right ) \frac{1}{2} \left ( 1+ \tanh\left\{{\frac { v-v_{\rm d}}{\frac{kT}{2e}}}\right\}\right ) \\
\frac{df}{dt} &=&  {\tau}^{-1}_{\rm Ca} \cosh\left\{{\frac {v-v_{\rm f}}{\frac{kT}{2e}}}\right\} \left[ \frac{1}{2}\left (1- \tanh\left\{{\frac { v-v_{\rm f}}{\frac{kT}{2e}}}\right\}\right )-f\right ] \;, \\
\end{eqnarray*}} \hfill
\parbox{1cm}{\begin{eqnarray}\label{eq37}\end{eqnarray}}

\noindent
and,

\parbox{14cm}{
\begin{eqnarray*}
i_{\rm Na} &=& g_{\rm Na}\,h\left (v-v_{\rm Na}\right ) \frac{1}{2} \left ( 1+ \tanh\left\{{\frac { v-v_{\rm m}}{\frac{kT}{2e}}}\right\}\right ) \\
\frac{dh}{dt} &=&  {\tau}^{-1}_{\rm Na} \cosh\left\{{\frac {v-v_{\rm h}}{\frac{kT}{2e}}}\right\} \left[ \frac{1}{2}\left (1- \tanh\left\{{\frac { v-v_{\rm h}}{\frac{kT}{2e}}}\right\}\right )-h\right ] \;,
\end{eqnarray*}} \hfill
\parbox{1cm}{\begin{eqnarray}\label{eq38}\end{eqnarray}}

\noindent
where $g_{\rm Ca}$ and $g_{\rm Na}$ are the conductances of the
calcium and sodium currents respectively, $v_{\rm Ca}$ and $v_{\rm
Na}$ are given by equations (\ref{eq12}) and (\ref{eq13}), $v_{\rm d}$
and $v_{\rm m}$ are the half--activation potentials, and $v_{\rm f}$
and $v_{\rm h}$ are the half--inactivation potentials.

\subsection{${\rm Na}^{+},{\rm K}^{+}$ Pump}
The Na,K--ATPase is found in the plasma membrane of virtually all
animal cells and is responsible for active transport of sodium and
potassium. Low sodium concentration and high potassium concentration
in the cytosol are essential for basic cellular functions such as
excitability, secondary active transport, and volume regulation. In
our model, the ${\rm Na}^{+},{\rm K}^{+}$ pump is the only energy
source.  We shall assume that the following equation is a complete
macroscopic description of the pump reaction (Chapman, 1978):

\begin{equation}
\label{eq48}
\begin{array}{ccc}
{\rm ATP} + 3 {\rm Na}_{\rm i}^{+} + 2 {\rm K}_{\rm e}^{+} & \stacking{{\LARGE \alpha}}{{\LARGE \beta}} {\rightleftharpoons} & {\rm ADP} + {\rm P}_{\rm io} + 3 {\rm Na}_{\rm e}^{+} + 2 {\rm K}_{\rm i}^{+} 
\end{array} \;,
\end{equation}

\noindent
where ATP, ADP and ${\rm P}_{\rm io}$ are adenosine triphosphate,
adenosine diphosophate and inorganic phosphate, while $\alpha$ and
$\beta$ are the rates of the forward and backward reactions. The energy 
involved in the movement of 3 ${\rm Na}^{+}$ and 2 ${\rm
K}^{+}$ ions against their electrochemical gradients is:

\begin{eqnarray}
\label{eq49}
{\Delta w}_{\rm Na} &=& - 3e (v-v_{\rm Na}) \\
\label{eq50}
{\Delta w}_{\rm  K} &=& + 2e (v-v_{\rm  K}) \;,
\end{eqnarray}

\noindent
where $v_{\rm K}$ and $v_{\rm Na}$ are given by equations (\ref{eq11}) and
(\ref{eq13}). This result is independant of the detailed interaction between ions, molecules
and the ATPase enzyme. Therefore, the total work done in reaction (\ref{eq48}) is:
\begin{eqnarray}
\Delta w &=& {\Delta w}_{\rm ATP} + {\Delta w}_{\rm Na} +{\Delta w}_{\rm K} \nonumber \\
\label{eq51}
         &=& {\Delta w}_{\rm ATP} + e(3v_{\rm Na} - 2v_{\rm  K} - v) \;,
\end{eqnarray}

\noindent
where ${\Delta w}_{\rm ATP}$ is the energy associated with the
breakdown of ATP. The ratio of $\alpha$ to $\beta$ is again given by
equations (\ref{eq30}) and (\ref{eq31}). The energy available from ATP
breakdown is much larger than the energy required to translocate the
potassium and sodium ions at small negative membrane potentials. In practice, 
such a pump or motorized swing door will soon reach saturation. We therefore 
choose the rate of the forward reaction to be constant, resembling the maximum
possible speed of the swing door in that direction, and obtain:
\begin{eqnarray}
\label{eq53}
\alpha &=& \lambda \\
\label{eq54}
\beta  &=& \lambda \exp\left[\frac{{\Delta w}_{\rm ATP} + e(3v_{\rm Na} - 2v_{\rm  K} - v)}{kT}\right] \;,
\end{eqnarray}
\noindent
where $\lambda$ is a constant. The net pump current for a cell with
$M$ pumps can then be written:
\begin{equation} 
\label{eq56}
i_{\rm NaK} = e M (\alpha - \beta) = k_{\rm NaK}\,\left ( 1-{\exp \left\{{\frac {-v -2v_{\rm K}+3v_{\rm Na}+v_{\rm ATP}}{\frac{kT}{e}}} \right\}}\right ) \nonumber \;,
\end{equation}

\noindent
where $k_{\rm NaK} = e\lambda M$, and $v_{\rm ATP} = {\Delta w}_{\rm
ATP}/e$.

\subsection{${\rm Na}^{+},{\rm Ca}^{2+}$ Exchanger}
To maintain a steady state for the intracellular calcium concentration
in for example heart cells, the amount of calcium that enters the cell
via ionic channels must be extruded. The ${\rm Na}^{+},{\rm
Ca}^{2+}$ exchanger is the major mechanism responsible for achieving a
balance between calcium entry and extrusion in oscillating cells. We
assume that the rates of the exchange reaction (Mullins, 1977) given
by:

\begin{equation}
\label{eq39}
\begin{array}{ccc}
3 {\rm Na}_{\rm e}^{+} + {\rm Ca}_{\rm i}^{2+} & \stacking{{\LARGE \alpha}}{{\LARGE \beta}} {\rightleftharpoons} & 3 {\rm Na}_{\rm i}^{+} + {\rm Ca}_{\rm e}^{2+} 
\end{array} \;,
\end{equation}

\noindent
are governed largely by the electrochemical gradients for sodium and
calcium, together with the membrane potential. In other words, the
energy produced when 3 extracellular sodium ions take the elevator
down into the cytosol is used to elevate one calcium ion up from the
cytosol into the extracellular space:
\begin{eqnarray}
\label{eq40}
{\Delta w}_{\rm Na} &=& + 3e (v-v_{\rm Na}) \\
\label{eq41}
{\Delta w}_{\rm Ca} &=& - 2e (v-v_{\rm Ca}) \;,
\end{eqnarray}

\noindent
where $v_{\rm Ca}$ and $v_{\rm Na}$ are given by equations (\ref{eq12}) and
(\ref{eq13}). The total work done in reaction (\ref{eq39}) is:
\begin{equation}
\label{eq42}
\Delta w = {\Delta w}_{\rm Na} +{\Delta w}_{\rm Ca} =  e (v - 3v_{\rm Na} + 2v_{\rm Ca}) \;.
\end{equation}

\noindent
The ratio of $\alpha$ to $\beta$ in equation (\ref{eq39}) is given by
equations (\ref{eq30}) and (\ref{eq31}). Since ${\Delta w} = e (v-3v_{\rm
Na} + 2v_{\rm Ca})$ in a cell usually oscillates around zero, the
most natural choice of $\alpha$ and $\beta$ is:
\begin{eqnarray}
\label{eq45}
\alpha &=& \lambda \exp\left[-\frac{e (v-3v_{\rm Na} + 2v_{\rm Ca})}{2kT}\right] \\
\label{eq46}
\beta  &=& \lambda \exp\left[+\frac{e (v-3v_{\rm Na} + 2v_{\rm Ca})}{2kT}\right] \;,
\end{eqnarray}

\noindent
\noindent
where again we make the {\em ad hoc} assumption that $\lambda$ is a
constant. For a cell with $N$ exchangers the net current is then:
\begin{equation}
\label{eq47}
i_{\rm NaCa} = - e N \left(\alpha - \beta \right) 
             = k_{\rm NaCa}\,\sinh\left\{{\frac {v- 3v_{\rm Na}+ 2v_{\rm Ca}}{\frac{2kT}{e}}}\right\} \;,
\end{equation}

\noindent
where $k_{\rm NaCa} = 2e\lambda N$.

\subsection{Membrane Voltage}
We assume that the electrical activity of a cell is described by the
five currents discussed above, and that all the other currents
(Boyett, 1996) are of minor importance. The standard differential
equation for the voltage, and the conservation laws for intracellular
ionic concentrations are then:
\begin{eqnarray}
\label{eq57}
\frac{dv}{dt} &=& -\frac{1}{C}
\left(i_{\rm K}+i_{\rm Ca}+i_{\rm Na}+i_{\rm NaCa}+i_{\rm NaK}
\right)\;,  \\
\label{eq58}
\frac{d}{dt}{[\rm K]_{\rm i}} &=& \frac {2i_{\rm NaK} - i_{\rm K}}{FV}\;, \\
\label{eq59}
\frac{d}{dt}{[\rm Ca]_{\rm i}} &=& \frac {2{i_{\rm NaCa}}- i_{\rm Ca}}{2FV}\;, \\
\label{eq60}
\frac{d}{dt}{[\rm Na]_{\rm i}} &=& \frac {-{i_{\rm Na}}- 3i_{\rm NaK}- 3i_{\rm NaCa}}{FV}\;,
\end{eqnarray}

\noindent
where $C$ is cell capacitance, $F$ is Faraday's constant and we assume
that the cell volume $V$ is constant. Solving equation (\ref{eq58})
for $i_{\rm K}$, equation (\ref{eq59}) for $i_{\rm Ca}$, and equation
(\ref{eq60}) for $i_{\rm Na}$, we obtain:
\begin{eqnarray}
\label{eq58b}
i_{\rm K} &=& -FV \frac{d}{dt}{[\rm K]_{\rm i}} + 2i_{\rm NaK} \;, \\
\label{eq59b}
i_{\rm Ca} &=& -2FV \frac{d}{dt}{[\rm Ca]_{\rm i}} + 2i_{\rm NaCa} \;, \\
\label{eq60b}
i_{\rm Na} &=& -FV \frac{d}{dt}{[\rm Na]_{\rm i}} - 3i_{\rm NaK}- 3i_{\rm NaCa} \;.
\end{eqnarray}

\noindent
Inserting this into equation (\ref{eq57}) we obtain:
\begin{equation}
\label{eq61}
\frac{dv}{dt} = \frac{FV}{C}\,\frac{d}{dt}
\left({[\rm K]_{\rm i}}+2{[\rm Ca]_{\rm i}}+{[\rm Na]_{\rm i}} \right)\;,
\end{equation}

\noindent
since all of the currents cancel. This equation can also be written as:

\begin{equation}
\label{eq61b}
\frac{d}{dt} \left(v - \frac{FV}{C} \left\{ [\rm K]_{\rm i} + 2[\rm Ca]_{\rm i} + [\rm Na]_{\rm i}  \right\} \right) = 0\;.
\end{equation}

\noindent
Integrating gives:
\begin{equation}
\label{eq62}
v - \frac{FV}{C}
\left( [\rm K]_{\rm i}+2[\rm Ca]_{\rm i}+[\rm Na]_{\rm i} \right) = c \;,
\end{equation}
\noindent
where the integration constant $c$ has to be determined. Given that the
voltage across a capacitor is zero when the net charge difference is zero,
choosing the integration constant as:
\begin{equation}
\label{eq63a}
c = -\frac{FV}{C} \left\{[\rm K]_{\rm e} + 2[\rm Ca]_{\rm e} + [\rm Na]_{\rm e} \right\} \;,
\end{equation}
\noindent
gives:

\begin{equation}
\label{eq63}
v = \frac{FV}{C} \left\{ ([\rm K]_{\rm i}-[\rm K]_{\rm e}) + 2([\rm Ca]_{\rm i}-[\rm Ca]_{\rm e}) + ([\rm Na]_{\rm i}-[\rm Na]_{\rm e}) \right\} \;,
\end{equation}
\noindent
so that the voltage is zero when there is no charge gradient. Equation
(\ref{eq63}) is simply the relation between the electric potential and
charge of a capacitor, which is in fact the origin of equation
(\ref{eq57}).  Thus it is completely general and independent of the
number of membrane currents in a model. The meaning of equation
(\ref{eq63}) is:
\begin{quote}
{\it The voltage across the membrane of a cell is caused by, and is
directly proportional to, the surplus of charge inside the cell,}
\end{quote}

\noindent
but it is only valid if we assume that the intracellular concentration
of anions is equal to the extracellular concentration of anions. Since
the cations are balanced by the anions in any solution, the
intracellular/extracellular concentration of anions must be close to
the intracellular/extracellular concentration of cations.

Since equation (\ref{eq63}) is the explicit integral of equation
(\ref{eq57}), it should be used instead of equation (\ref{eq57}) (or
the equivalent of equation (\ref{eq57})) in any model. The
differential equation (\ref{eq57}) is needed only in models
where the intracellular ionic concentrations are not tracked
individually (for example, the Hodgkin--Huxley equations (1952)).

There is a significant difference between equations (\ref{eq57}) and
(\ref{eq63}). Here is an analogy: There are two different ways to
calculate how many ions are inside a cell. The first method counts
every ion enetering or leaving (equation (\ref{eq57})), while the
second method simultaneously counts all the ions inside the cell,
perhaps by means of an ion detector (equation (\ref{eq63})). Both
methods will correctly give the {\em variation} in the number of ions
in the cell. However, the observer of ions entering and leaving will
only observe {\it variations} in the number of ions.  If he wants to
know the actual number of ions in the cell, he must make an initial
guess of the number of ions already inside. Because his guess may
differ significantly from the actual ion number (as maesured by the
ion detector) indicates, the results from the two methods may be
contradictory.

\subsection{Osmotic Pressure}
In this section we demonstrate that variations in the osmotic pressure
are a natural consequence of our model. In fact, van't Hoff's equation for
the osmotic pressure can be derived by tracking the flow of energy
under the assumption of constant temperature and volume.

From elementary physics the work $W$ done by a force ${\cal F}$ to
move an object from ${\cal X=A}$ to ${\cal X=B}$ is given by:
\begin{equation}
\label{eq68a}
W = \int_{\cal A}^{\cal B} {\cal F} d {\cal X} = {\cal F}({\cal B}-{\cal A}) = \int_{0}^{t} {\cal F} \frac{d {\cal X}}{dt} dt = \int_{0}^{t} {\cal F} {\cal V} dt \,,
\end{equation}

\noindent
where ${\cal V} = \frac{d {\cal X}}{dt}$ is the velocity of the
object, and ${\cal X}(0)={\cal A}$, ${\cal X}(t)={\cal B}$.  When ions
move across the membrane of a cell the transmembrane voltage $v$ is
like a force and the current $i$ is like a velocity, thus:

\begin{equation}
\label{eq68}
W = \int_{0}^{t} v i dt \,.
\end{equation}

\noindent
If we take into account the reversal potentials in our model
this can be written:

\begin{eqnarray}
\label{eq69}
W &=& \int_{0}^{t} i_{\rm K}(v-v_{\rm K}) + i_{\rm Ca}(v-v_{\rm Ca}) + i_{\rm Na}(v-v_{\rm Na}) \nonumber \\
  &+& i_{\rm NaCa}(v- 3v_{\rm Na}+ 2v_{\rm Ca}) + i_{\rm NaK}(v +2v_{\rm K}-3v_{\rm Na}) dt \;.
\end{eqnarray}

\noindent
Using equations (\ref{eq57}), (\ref{eq58b}), (\ref{eq59b}) and
(\ref{eq60b}) the currents cancel and we obtain:
\begin{equation}
\label{eq70}
W = -C \int_{0}^{v} vdv + FV \int_{[\rm K]_{\rm e}}^{[\rm K]_{\rm i}} v_{\rm K} d([{\rm K}]_{\rm i}) + 2FV \int_{[\rm Ca]_{\rm e}}^{[\rm Ca]_{\rm i}} v_{\rm Ca} d([{\rm Ca}]_{\rm i}) + FV \int_{[\rm Na]_{\rm e}}^{[\rm Na]_{\rm i}} v_{\rm Na} d([\rm Na]_{\rm i})
\end{equation}

\noindent
When solving these integrals it is important to remember that the
reversal potentials $v_{\rm K}$, $v_{\rm Ca}$ and $v_{\rm Na}$ given
by equations (\ref{eq11}), (\ref{eq12}) and (\ref{eq13}), are
dependent on the integration variables $[\rm K]_{\rm i}$, $[\rm
Ca]_{\rm i}$ and $[\rm Na]_{\rm i}$. Using:
\begin{equation}
\label{eq71}
\int \ln(\phi) d\phi = \phi \ln(\phi) - \phi \;,
\end{equation}

\noindent
the integral becomes:

\parbox{14cm}{
\begin{eqnarray*}
W = -\frac{1}{2} C v^2 &-& RTV \left\{ [\rm K]_{\rm i}\, \ln ({\frac {[\rm K]_{\rm i}}{[\rm K]_{\rm e}}}) + [\rm Ca]_{\rm i} \,\ln ({\frac {[\rm Ca]_{\rm i}}{[\rm Ca]_{\rm e}}}) + [\rm Na]_{\rm i}\,\ln ({\frac {[\rm Na]_{\rm i}}{[\rm Na]_{\rm e}}}) \right\} \\
&-& RTV \left\{ [\rm K]_{\rm e}-[\rm K]_{\rm i}+[\rm Na]_{\rm e}-[\rm Na]_{\rm i}+[\rm Ca]_{\rm e}-[\rm Ca]_{\rm i} \right\} \,.
\end{eqnarray*}} \hfill
\parbox{1cm}{\begin{eqnarray}\label{eq72}\end{eqnarray}}

\noindent
This is the total work associated with the currents. The pump work and
work due to the exchanger and channel currents have opposite signs
since the pump moves ions against their electrochemical gradients,
while energy is lost by the downhill ionic fluxes through the
exchanger and the channels. Since energy cannot be created or
destroyed, we define the potential energy $P$ of the cell so that the
total energy $E$ is zero (this is the most common definition of
potential energy):
\begin{equation}
\label{eq666}
E = P + W = 0 \;,
\end{equation}

\noindent
thus,

\parbox{14cm}{
\begin{eqnarray*}
P = \frac{1}{2} C v^2 &+&  RTV \left\{ [\rm K]_{\rm i}\, \ln ({\frac {[\rm K]_{\rm i}}{[\rm K]_{\rm e}}}) + [\rm Ca]_{\rm i} \,\ln ({\frac {[\rm Ca]_{\rm i}}{[\rm Ca]_{\rm e}}}) + [\rm Na]_{\rm i}\,\ln ({\frac {[\rm Na]_{\rm i}}{[\rm Na]_{\rm e}}}) \right\} \\
&+& RTV \left\{ [\rm K]_{\rm e}-[\rm K]_{\rm i}+[\rm Na]_{\rm e}-[\rm Na]_{\rm i}+[\rm Ca]_{\rm e}-[\rm Ca]_{\rm i} \right\} \,.
\end{eqnarray*}} \hfill
\parbox{1cm}{\begin{eqnarray}\label{eq67}\end{eqnarray}} 

\noindent
The first term is the electrical potential energy of a capacitor, the
second term is the potential energy associated with the concentration
gradients, and the third term is the potential energy associated with
an osmotic pressure gradient (caused by the concentration
gradients). Rewriting equation (\ref{eq666}):

\begin{eqnarray}
\label{eq101}
\int_{0}^{t} i_{\rm NaK}(v &+& 2v_{\rm K} - 3v_{\rm Na}) dt =\\
\label{eq102}
&-& \frac{1}{2} C v^2 \\
\label{eq103}
&-& RTV \left\{ [\rm K]_{\rm i}\, \ln ({\frac {[\rm K]_{\rm i}}{[\rm K]_{\rm e}}}) + [\rm Ca]_{\rm i} \,\ln ({\frac {[\rm Ca]_{\rm i}}{[\rm Ca]_{\rm e}}}) + [\rm Na]_{\rm i}\,\ln ({\frac {[\rm Na]_{\rm i}}{[\rm Na]_{\rm e}}}) \right\} \\
\label{eq104}
&-& RTV \left\{ [\rm K]_{\rm e}-[\rm K]_{\rm i}+[\rm Na]_{\rm e}-[\rm Na]_{\rm i}+[\rm Ca]_{\rm e}-[\rm Ca]_{\rm i} \right\} \\
\label{eq105}
&-& \int_{0}^{t} i_{\rm K}(v-v_{\rm K}) + i_{\rm Ca}(v-v_{\rm Ca}) + i_{\rm Na}(v-v_{\rm Na}) + i_{\rm NaCa}(v- 3v_{\rm Na}+ 2v_{\rm Ca}) dt \;,
\end{eqnarray}

\noindent
we see that the pump (equation (\ref{eq101})) produces:

\begin{enumerate}
\item a transmembrane voltage gradient (equation (\ref{eq102})),
\item a transmembrane concentration gradient (equation (\ref{eq103})),
\item a transmembrane pressure gradient (equation (\ref{eq104})), and 
\item downhill fluxes through the exchanger and the channels (equation (\ref{eq105})).
\end{enumerate}

In 1886 van't Hoff noticed that the behaviour of solutes in dilute
solutions resembles the behaviour of a perfect gas:

\begin{quote}
{\footnotesize The pressure which a gas exerts when a given number of
molecules are distributed in a given volume is equally great as the
osmotic pressure, which under the same conditions would be produced my
most solutes when they are dissolved in an arbitrary solvent.}
\end{quote}

\noindent
Therefore we can use the ideal gas law $p = [S] RT$ to calulate the
osmotic pressure across the cell membrane:

\begin{equation}
\label{eq1000}
\pi = RT \left\{ [\rm K]_{\rm i}-[\rm K]_{\rm e}+[\rm Na]_{\rm i}-[\rm Na]_{\rm e}+[\rm Ca]_{\rm i}-[\rm Ca]_{\rm e} \right\} \;.
\end{equation}

\noindent
This is the van't Hoff equation for the osmotic pressure ($\pi$)
across a solute impermeable barrier separating two ideal dilute
solutions. This is nothing but equation (\ref{eq104}) divided by the
cell volume $V$. 

Since we did not have the osmotic pressure in mind when we made the
theory, our theory is a perfect example of one of Richard P. Feynman's
astute observations:

\begin{quote}
{\footnotesize When you have put a lot of ideas together to make an
elaborate theory, you want to make sure, when explaining what it fits,
that those things it fits are not just the things that gave you the
idea of the theory; but that the finished theory makes something else
come out right, in addition.}
\end{quote}

\noindent
Our theory indeed has ``something else [that] comes out right'' -- the
osmotic pressure variations.

\section{Model Parameters}
A mathematical model of the membrane potential has been derived;
equations (\ref{eq11}), (\ref{eq12}) and (\ref{eq13}) represent the
equilibrium potentials, equations (\ref{eq36}), (\ref{eq37}) and
(\ref{eq38}) the ionic currents, equations (\ref{eq47}) and
(\ref{eq56}) the exchanger and the pump currents, equations
(\ref{eq58}), (\ref{eq59}) and (\ref{eq60}) the ionic concentrations,
equation (\ref{eq63}) the membrane voltage, and finally, equation
(\ref{eq1000}) the osmotic pressure across the cell membrane. The
model has 6 dimensions with the variables $x$, $f$, $h$, $[\rm K]_{\rm
i}$, $[\rm Ca]_{\rm i}$ and $[\rm Na]_{\rm i}$. 
\begin{figure}
\epsfxsize=0.95\textwidth
\epsffile{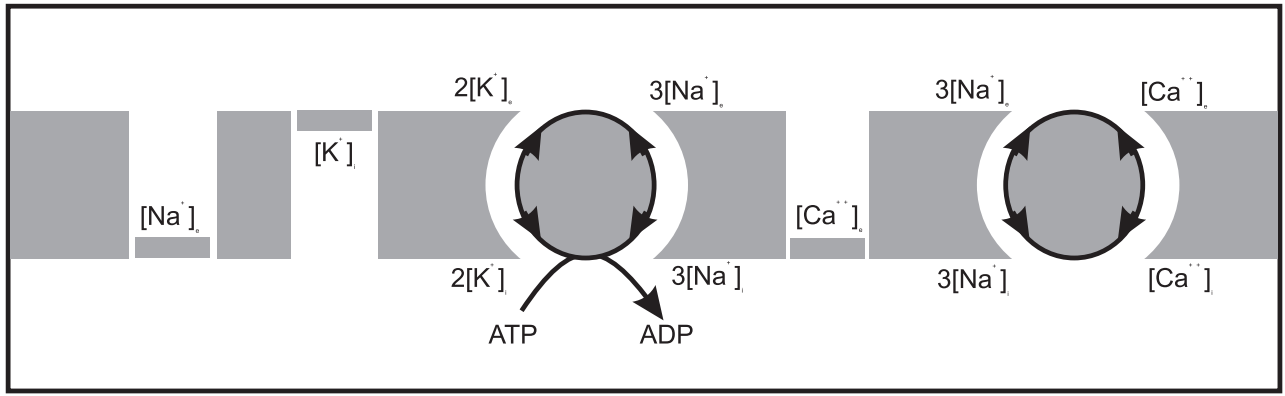}
\vspace{0.5cm}
\caption{Schematic diagram of the simplified cell model. The three
ionic currents of potassium, calcium and sodium, the exchanger and the
pump. The ``moveable'' gating doors of the ionic channels illustrate
that the concentration gradient of potassium is the opposite to the
concentration gradients of calcium and sodium.}
\end{figure}

Fig. 1 shows a schematic diagram of the model, with the ionic channels
of potassium, calcium and sodium, the ${\rm Na}^{+},{\rm Ca}^{2+}$
exchanger and the ${\rm Na}^{+},{\rm K}^{+}$ pump. The ``moveable''
gating doors of the ionic channels illustrate that the concentration
gradient of potassium is the opposite of the concentration gradients
of calcium and sodium.

On the basis of distinct biophysical and pharmacological properties,
cardiac calcium currents have been classified into a long lasting type
($i_{{\rm Ca,L}}$) and a transient type ($i_{{\rm Ca,T}}$), while
cardiac delayed rectifier potassium currents have been classified into
a rapid type ($i_{{\rm Kr}}$) and a slow type ($i_{{\rm Ks}}$). In our
model we assume that $i_{{\rm Ca,T}}$ and $i_{{\rm Ks}}$ are of minor
importance; i.e. when we talk about $i_{\rm Ca}$ we mean $i_{{\rm
Ca,L}}$, and when we talk about $i_{\rm K}$ we mean $i_{{\rm Kr}}$.

First, we want to justify the presence of the term $kT/2e$ in equations
(\ref{eq36}), (\ref{eq37}) and (\ref{eq38}). This corresponds to a
slope factor for the activation and inactivation curves of $kT/4e
\approx 6.68 \, {\rm mV}$ at $37^{\circ} {\rm C}$. The observed slope
factors are $7.4 \, {\rm mV}$ for activation of $i_{\rm K}$
(Shibasaki, 1987), $6.6 \, {\rm mV}$ for activation of $i_{\rm Ca}$
(Hagiwara {\em et al.}, 1988), $6.0 \, {\rm mV}$ for inactivation of
$i_{\rm Ca}$ (Hagiwara {\em et al.}, 1988), $6.0 \, {\rm mV}$ for
activation of $i_{\rm Na}$ (Muramatsu {\em et al.}, 1996), and,
finally, $6.4 \, {\rm mV}$ for inactivation of $i_{\rm Na}$ (Muramatsu
{\em et al.}, 1996). Hence, we see that $kT/4e$, corresponding to a
gating charge of $q \approx 4e$, is an excellent approximation. Now,
we would like to distinguish between the fundamental physical
constants (table \ref{table1}), the experimentally observed constants
(table \ref{table2}), the adjustable parameters (table \ref{table3})
and the initial conditions (table \ref{table4}) in the model.
\vbox{\begin{table}[h]
\caption{\centerline{Fundamental Physical Constants}}
\begin{tabular}{lrc} 
\multicolumn{1}{c}{Name} &
\multicolumn{1}{c}{Value} &
\multicolumn{1}{c}{Unit} \\
\tableline
$k              $ & $  1.38065812 \cdot  10^{-20}     $ & mJ/K \\
$e              $ & $  1.6021773349 \cdot  10^{-19}   $ & C \\
$F              $ & $  96485.30929                    $ & C/mol \\
$T              $ & $  310.15                         $ & K  \\
$R=kF/e         $ & $  8314.511935                    $ & J/kmol K
\end{tabular}
\label{table1}
\end{table}}

The half--activation and inactivation potentials in the model ($v_{\rm
x}$, $v_{\rm d}$, $v_{\rm f}$, $v_{\rm m}$ and $v_{\rm h}$) are based
on the experiments of Shibasaki (1987), Hagiwara {\em et al.} (1988)
and Muramatsu {\em et al.} (1996), and we use a value of $v_{\rm ATP}$
that gives a reversal potential for the sodium pump in good agreement
with the experiments of Sakai {\em et al.} (1996). The maximum time
constants in these experiments were $ 203 \, {\rm ms}$ for activation
of $i_{\rm K}$ (Shibasaki, 1987), $225 \, {\rm ms}$ for inactivation
of $i_{\rm Ca}$ (Hagiwara {\em et al.}, 1988) and $174 \, {\rm ms}$
for inactivation of $i_{\rm Na}$ (Muramatsu {\em et al.}, 1996). In
the model, however, we combine these and use a maximum time constant of
$200 \, {\rm ms}$ for both ${\tau}_{\rm K}$, ${\tau}_{\rm Ca}$ and
${\tau}_{\rm Na}$. Finally, we use typical values for cell volume, cell
capacitance, and extracellular ionic concentrations.
\vbox{\begin{table}[h]
\caption{\centerline{Observed Constants}}
\begin{tabular}{lrc} 
\multicolumn{1}{c}{Name} &
\multicolumn{1}{c}{Value} &
\multicolumn{1}{c}{Unit} \\
\tableline
$[\rm K]_{e}              $ & $          5.4  $ & mM \\
$[\rm Ca]_{e}             $ & $            2  $ & mM \\
$[\rm Na]_{e}             $ & $          140  $ & mM \\
$V                        $ & $           10  $ & ${\rm 10^3 \mu m^3}$ \\
$C                        $ & $           47  $ & pF \\
$v_{\rm x}                $ & $        -25.1  $ & mV \\
$v_{\rm d}                $ & $         -6.6  $ & mV \\
$v_{\rm f}                $ & $        -25.0  $ & mV \\
$v_{\rm m}                $ & $        -41.4  $ & mV \\
$v_{\rm h}                $ & $        -91.0  $ & mV \\
$v_{\rm ATP}              $ & $         -450  $ & mV \\
$\tau                     $ & $          200  $ & ms \\
\end{tabular}
\label{table2}
\end{table}}

The density of ionic channels, exchangers and pumps (i.e. $g_{\rm
Ca}$, $g_{\rm Na}$, $g_{\rm K}$, $k_{\rm NaK}$ and $k_{\rm NaCa}$) can
vary significantly from cell to cell. In order to reproduce the action
potentials recorded in a spontaneously beating adult rabbit sinoatrial
node cell with normal Tyrode solution as external super fusing
solution (Fig. 7 A. in Baruscotti {\em et al.} (1996)), we fit the
adjustable parameters (table \ref{table3}) and the initial conditions
(table \ref{table4}) numerically. More details are given in (Endresen,
1997a). Many combinations of $g_{\rm Ca}$, $g_{\rm Na}$, $g_{\rm K}$,
$k_{\rm NaK}$ and $k_{\rm NaCa}$ resulted in good approximations to
the experimentally recorded waveform, from which we conclude that
different cells can produce the same action potential although they
have a different mixture of ionic channels, exchangers and pumps. The
final set of parameters presented in table \ref{table3} are based on
the choice $k_{\rm NaK} = 12.2 \, {\rm pA}$.
\vbox{\begin{table}[h]
\caption{\centerline{Adjustable Parameters}}
\begin{tabular}{lrc} 
\multicolumn{1}{c}{Name} &
\multicolumn{1}{c}{Value} &
\multicolumn{1}{c}{Unit} \\
\tableline
$g_{\rm Ca}               $ & $      9.29045 $ & nS \\
$g_{\rm Na}               $ & $    253.94203 $ & nS \\
$g_{\rm K}                $ & $      0.70302 $ & nS \\
$k_{\rm NaCa}             $ & $   8181.31568 $ & pA \\
$k_{\rm NaK}              $ & $     12.20000 $ & pA 
\end{tabular}
\label{table3}
\end{table}}

\vbox{\begin{table}[h]
\caption{\centerline{Initial Conditions}}
\begin{tabular}{lrc} 
\multicolumn{1}{c}{Name} &
\multicolumn{1}{c}{Value} &
\multicolumn{1}{c}{Unit} \\
\tableline
$x_{\rm 0}                         $ & $                   0 $ & -- \\
$f_{\rm 0}                         $ & $                   1 $ & -- \\
$h_{\rm 0}                         $ & $                   0 $ & -- \\
${[\rm K]_{\rm i}}_{\rm 0}         $ & $          130.880955 $ & mM \\
${[\rm Ca]_{\rm i}}_{\rm 0}        $ & $            0.000790 $ & mM \\
${[\rm Na]_{\rm i}}_{\rm 0}        $ & $           18.514880 $ & mM 
\end{tabular}
\label{table4}
\end{table}}

\newpage
\section{Simulation Results}
The six differential equations in the model were solved numerically
using a fifth--order Runge--Kutta method with variable steplength,
more details are given in (Endresen, 1997b). Fig. 2 shows the five
membrane currents and Fig. 3 shows the recorded and simulated action
potentiald together with the intracellular ionic concentrations. These
computations used the initial conditions stated in table \ref{table4}.

Cells must generate their membrane potential by actively transporting
ions against their concentration gradients. To examine this process in
our model, we ran a simulation starting with equal intracellular and
extracellular ionic concentrations:
\begin{eqnarray}
{[\rm K]_{\rm i}}  &=& 5.4 \, {\rm mM} \nonumber \;, \\
\label{eq73}
{[\rm Ca]_{\rm i}} &=& 2 \, {\rm mM} \;, \\
{[\rm Na]_{\rm i}} &=& 140 \, {\rm mM} \nonumber \;.
\end{eqnarray} 

\ \\
\vspace{6.5cm}
\ \\
\vbox{\begin{figure}[h]
\label{fig2}
\epsfxsize=0.15\textwidth
\hspace{-2.0cm} \epsffile{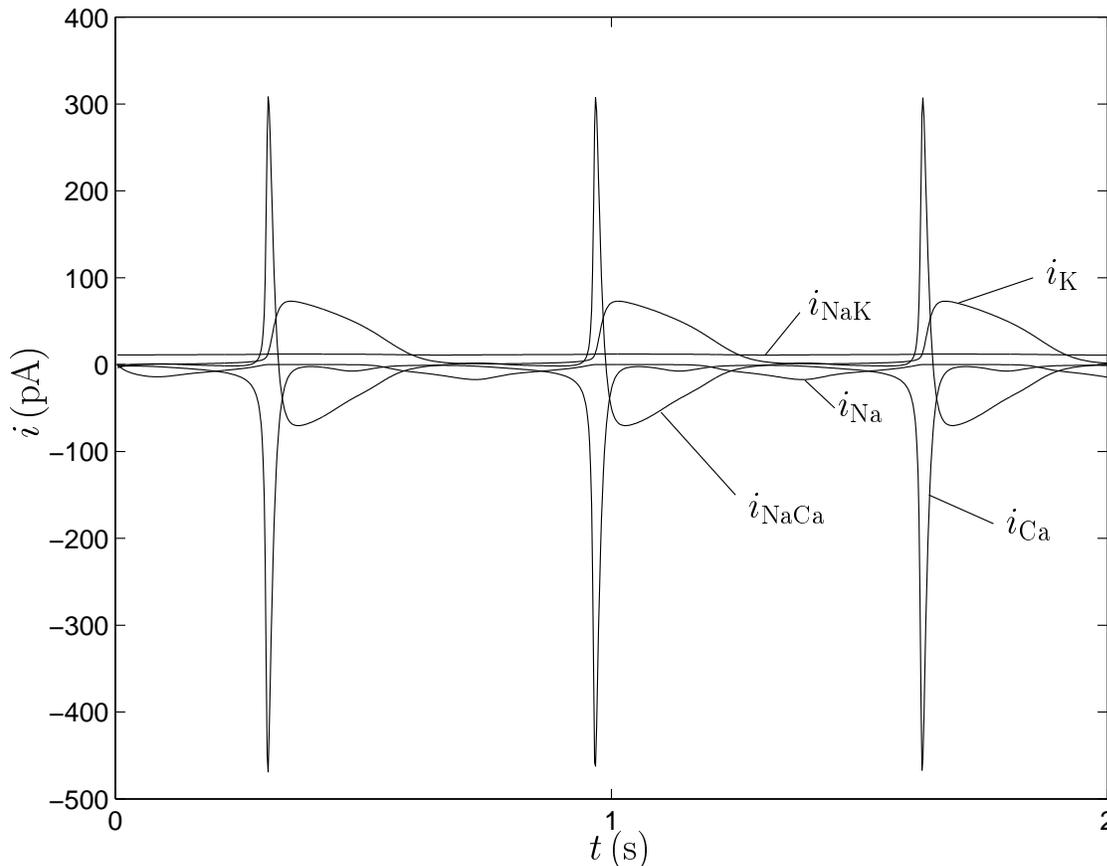}
\caption{Membrane currents in a simple ionic model of a rabbit
sinoatrial node cell. We show the outward delayed rectifying potassium
current ($i_{\rm K}$), the inward calcium current ($i_{\rm Ca}$), the
inward sodium current ($i_{\rm Na}$), the sodium calcium exchange
current ($i_{\rm NaCa}$) and the sodium potassium pump current
($i_{\rm NaK}$). These computations used the initial conditions in
table \ref{table4}. }
\end{figure}}

\noindent
The results are presented in Fig. 4. After approximately 1500 seconds
(25 minutes) the system appears to reach a stable fixed point with the
following intracellular ionic concentrations:
\begin{eqnarray}
{[\rm K]_{\rm i}}  &=& 115.842881 \, {\rm mM} \nonumber \;, \\
\label{eq74}
{[\rm Ca]_{\rm i}} &=& 4.485016 \cdot 10 ^{-5} \, {\rm mM} \;, \\
{[\rm Na]_{\rm i}} &=& 33.548671 \, {\rm mM} \nonumber \;.
\end{eqnarray} 

\noindent
At 2500 seconds we kick the cell with a $20 \, {\rm pA}$ pulse of
potassium ions with duration $50 \, {\rm ms}$ to see if it wants to
oscillate. It surely does, and the oscillations soon become
almost identical to the original oscillations present in Fig. 2 and 3
(you can not see this from Fig. 4 since the time scale is much to
big). The values for the concentrations only differ slightly from the
initial conditions in table \ref{table4}, as indicated by the final
concentrations calculated by the program (at 5000 seconds):
\begin{eqnarray}
{[\rm K]_{\rm i}}  &=& 131.075490 \, {\rm mM} \nonumber \;, \\
\label{eq75}
{[\rm Ca]_{\rm i}} &=& 6.827191 \cdot 10 ^{-4} \, {\rm mM} \;, \\
{[\rm Na]_{\rm i}} &=& 18.320693 \, {\rm mM} \nonumber \;.
\end{eqnarray} 
\ \\
\vspace{4cm}
\ \\
\vbox{\begin{figure}[h]
\label{fig3}
\epsfxsize=0.16\textwidth
\hspace{-2.0cm} \epsffile{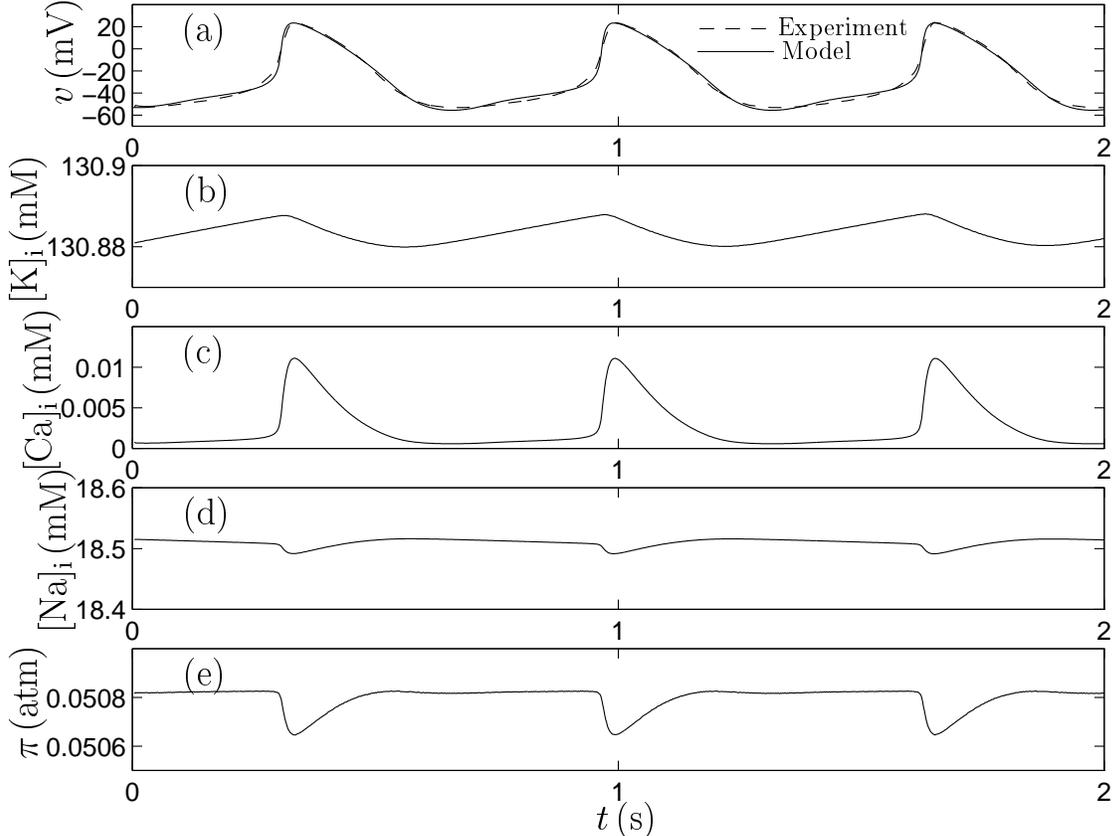}
\caption{Membrane potential, intracellular ionic concentrations and
osmotic pressure of a rabbit sinoatrial node cell. (a)
Model--generated (solid line) and experimentally recorded (dashed
line) action potential waveform, (b) potassium concentration $[\rm
K]_{i}$, (c) calcium concentration $[\rm Ca]_{i}$, (d) sodium
concentration $[\rm Na]_{i}$ and (e) the osmotic pressure $\pi$ across
the cell membrane.  These computations used the initial conditions
in table \ref{table4}.}
\end{figure}}

\noindent
The trajectory reaches the limit cycle at 2560 seconds, and there is
no drift in the intracellular ionic concentrations from this point to
the end of the simulation (at 5000 seconds). The long time simulation
in Fig. 4 is a numerical indication that the oscillation in Fig. 2 and
3 is indeed a stable limit cycle. However, the model also has a stable
fixed point given by (\ref{eq74}).

\section{Discussion} 
We have presented a simple model for the cells of the rabbit
sinoatrial node. Our model involves only ${\rm Na}^{+},{\rm K}^{+}$,
and ${\rm Ca}^{2+}$ ions, their respective channels, the ${\rm
Na}^{+},{\rm Ca}^{2+}$ exchanger, and the ${\rm Na}^{+},{\rm K}^{+}$
pump. The equations were derived using basic physical principles and
conservation laws. Since the only source of energy in our model is the
sodium pump, we could easily track the flow of energy in our
system. We showed that the pump works to generate a transmembrane
voltage, pressure gradient and ionic concentration gradients.  Our
equations also accounted for the energy lost due to downhill ionic
fluxes through the exchanger and channels. A prediction of osmotic
pressure variations was a novel result of our energy analysis.
 
The intracellular ionic concentrations are dynamic variables in our
model, governed by the conservation equations (\ref{eq58}),
(\ref{eq59}), and (\ref{eq60}). This allowed us to replace the
standard differential equation for the voltage (\ref{eq57}) with the
algebraic equation (\ref{eq63}). Although a number of other ionic
models also keep track of intracellular ionic concentrations (see
Wilders (1993)), we are unaware of any other model using an algebraic
equation for the membrane potential. Models that use the standard
voltage differential equation (\ref{eq57}) have a superfluous extra
dimension. Furthermore, the initial conditions for this extra
differential equation cannot be chosen independently of the initial
conditions of the conservation equations (\ref{eq58}), (\ref{eq59}),
and (\ref{eq60}) -- otherwise, the computed membrane potential will be
erroneous. For these reasons, we suggest that our algebraic expression
for the membrane potential should replace the standard voltage
differential equation in models where intracellular ionic
concentrations are dynamic variables.

\ \\
\vspace{5.0cm}
\ \\
\vbox{\begin{figure}[h]
\label{fig4}
\epsfxsize=0.21\textwidth
\hspace{-3.0cm} \epsffile{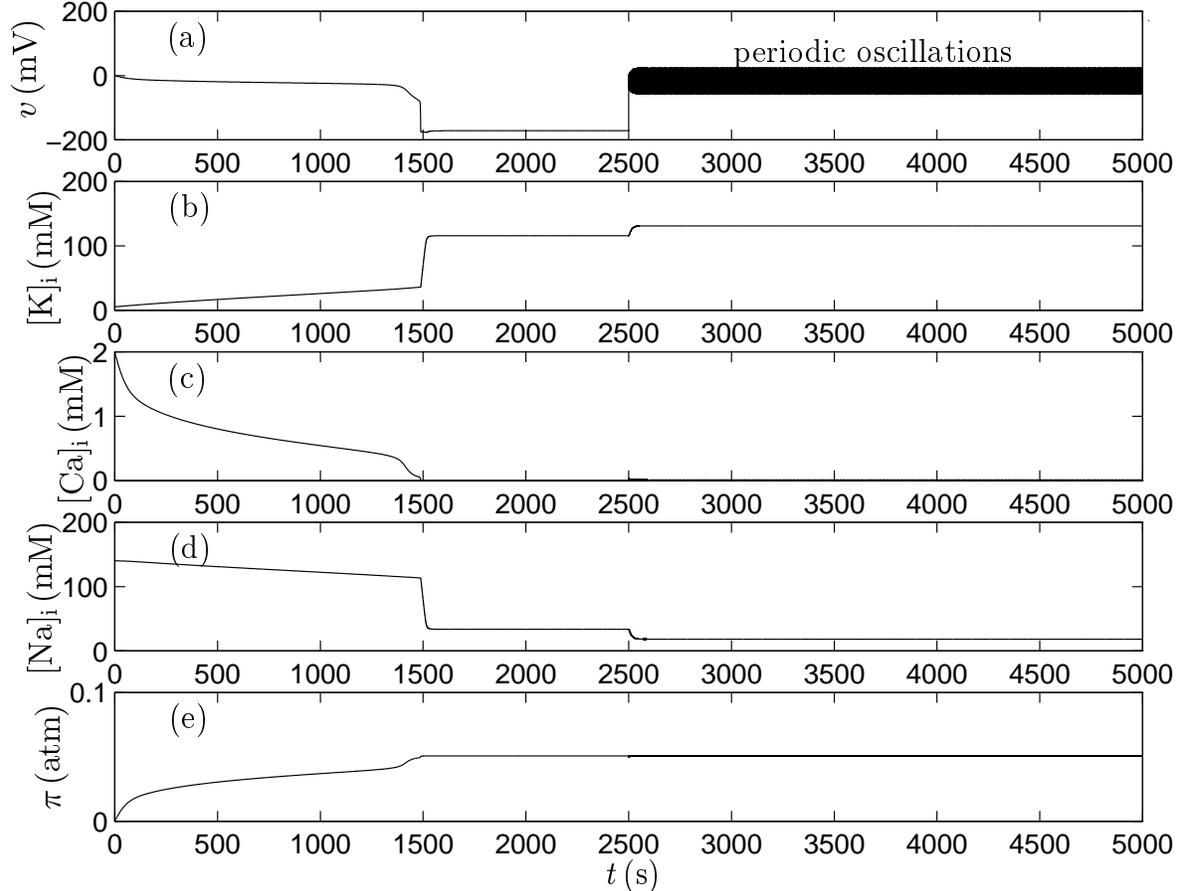}
\caption{Long time simulation showing the membrane potential,
intracellular ionic concentrations and osmotic pressure starting with
equal intracellular and extracellular concentrations: ${[\rm K]_{\rm
i}} = 5.4 \, {\rm mM}$, ${[\rm Ca]_{\rm i}} = 2 \, {\rm mM} $ and
${[\rm Na]_{\rm i}} = 140 \, {\rm mM} $. The cell is given a $20 \,
{\rm pA}$ pulse of potassium with duration $50 \, {\rm ms}$ after 2500
seconds of the simulation. (a) Membrane potential $v$, (b) potassium
concentration $[\rm K]_{i}$, (c) calcium concentration $[\rm Ca]_{i}$,
(d) sodium concentration $[\rm Na]_{i}$ and (e) the osmotic pressure
$\pi$ across the cell membrane.}
\end{figure}}

Our model does not include funny current ($i_{\rm f}$), ATP sensitive
channels, stretch-activated channels, or other ion channels that may
be important (Boyett, 1996). We also ignored the effect of calcium
uptake and release from the sarcoplasmatic reticulum, and the effect
of anions like chloride. We have assumed that the ionic channels are
governed by a Markov process. However, since transmembrane proteins
can have a large number of conformational states (Liebovitch, 1995),
perhaps a fractal model for the gating would be a better approach. We
assumed that the maximum of the activation/inactivation time constant
occurs at the same voltage as the inflection point of the sigmoidal
steady state activation/inactivation curve. Also, we have assumed that
the cell volume is constant. While such assumptions reduce the number
of parameters in the model, they may also result in discrepancies with
experiment. A natural extension of our model would include a variable
cell volume.

Finally, we would like to point out that our model is based on
experiments where some were conducted at room temperature
(22--$24^{\circ} {\rm C}$) (Baruscotti {\em et al.}, 1996; Muramatsu
{\em et al.}, 1996), while others were performed at $37^{\circ} {\rm
C}$ (Shibasaki, 1987; Hagiwara {\em et al.}, 1988; Sakai {\em et al.},
1996). It is not clear what affect this inconsistency had on the
accuracy of our model.
 
The values of the parameters $g_{\rm Ca}$, $g_{\rm Na}$, $g_{\rm K}$,
$k_{\rm NaK}$ and $k_{\rm NaCa}$, given in table \ref{table3}, are
only an estimate of the actual physiological parameters. We did not
systematically study the dynamics of the model for different
parameters. However, interested readers can download the source code
for the model (in both UNIX and Windows NT environments) from the
following internet site: 
\begin{quote}
\hspace{2cm} http://www.physio.mcgill.ca/guevaralab/singlecell.htm
\end{quote} 
\noindent 
We hope that future experiments will help us to discriminate between the  
different parameter sets that reproduce the experimentally recorded 
action potentials. 

\acknowledgements 
\noindent
Lars Petter Endresen would like to thank professor Jan Myrheim for his
wonderful way of explaining classical physics, and professor Per Jynge
for giving a fascinating introduction to the exciting field of cardiac
electrophysiology. Discussions with Per Hemmer, Johan Skule H{\o}ye,
K{\aa}re Olaussen, Michael Guevara, Michael Mackey and Jacques Belair
have been essential. Nils Skarland and Caroline Chopra deserve credit
for improving the manuscript. Special thanks to Aoxiang Xu for
testing, verifying and implementing the model on a Windows NT
Workstation, and to Ali Yehia for explaining the cited experimental
papers to a theorist. Lars Petter Endresen was supported by a
fellowship at NTNU, and has received support from The Research Council
of Norway (Programme for Supercomputing) through a grant of computing
time. Kevin Hall receives support from the Medical Research Council of
Canada.


\end{document}